\documentclass{iopjournal}

\usepackage{amsmath}
\usepackage{amssymb}

\usepackage[normalem]{ulem}

\begin{document}

\articletype{Article type} %

\title{How Phase Coexistence affects the mechanical properties of heterogeneous 2D suspensions}

\author{Raúl Molina-Prados Lallena$^{1,2}$, José Martín-Roca$^{3,*}$ , Kristian Thĳssen$^{4}$, Tyler Shendruk$^{4}$ , Angelo Cacciuto$^{5,*}$ and Chantal Valeriani$^{1,2,*}$}

\affil{$^1$Departamento de Electrónica, Física Átomica y Térmica, Universidad Complutense de Madrid, Madrid, Spain.}

\affil{$^2$GISC, Madrid}

\affil{$^3$ Departament de Física  de la Materia Condensada, Universitat de Barcelona, C. Martı Franques 1, 08028 Barcelona, Spain}

\affil{$^4$Niels Bohr Institute, University of Copenhagen, Blegdamsvej 17, Copenhagen, Denmark.}

\affil{$^5$School of Physics and Astronomy, The University of Edinburgh, Peter Guthrie Tait Road, Edinburgh, EH9 3FD, United Kingdom}

\affil{$^3$Department of Chemistry, Columbia University, New York, NY, 10027, USA.}

\affil{$^*$Author to whom any correspondence should be addressed.}

\email{josema10@ucm.es,cvaleriani@ucm.es,ac2822@columbia.edu }


\begin{abstract}
Although numerical simulations of rheological measurements typically focus on homogeneous systems, heterogeneity can profoundly impact material properties. 
We report on the rheological properties of a suspension of two-dimensional Lennard-Jones particles across the gas/liquid and the gas/solid coexistence lines of the system. We show how the presence of multiple coexisting states has a significant impact on the mechanical properties of these systems when compared with their homogeneous reference counterparts. Our results establish an extended map to navigate a landscape where not only density and temperature, but also phase coexistence, dictate the transition from viscous to elastic-dominated behavior under shear. These results provides a benchmark for future research into 
heterogeneous fluids where the coexistence of complex dynamic states is frequently observed.
\end{abstract}

\section{Introduction}

Studying the  rheology of particle suspensions is crucial for understanding their mechanical properties and assess their potential applications in materials engineering~\cite{flenner2019viscoelastic,Koumakis2015}. 
While numerous studies have reported on the rheology of colloidal suspensions in liquid, crystalline, or glassy states~\cite{flenner2019viscoelastic,heyes1986shear,larson1999structure}, these approaches largely rely on the assumption of spatial uniformity. 
Recent work has demonstrated that intrinsically out-of-equilibrium, active dynamic heterogeneities can give rise to nontrivial couplings between microscopic structure and macroscopic dynamics~\cite{martin2025motility,sheer2025,mo2024,Morse2021}. 
Surprisingly, little is known about how analogous heterogeneous effects manifest in thermalised systems, where equilibrium constraints fundamentally alter stress transmission and relaxation pathways.
A striking example of such heterogeneity arises in suspensions of attractive particles under thermodynamic conditions of gas/liquid or gas/solid phase coexistence~\cite{smit1991vapor,hansen1969phase,li2020LJ2D,vanMeel2008,Trokhymchuk1999,Stephan2018}.

Historically, the two-dimensional Lennard-Jones (LJ) fluid model has served as a prototype for many systems~\cite{barker1981phase, smit1991vapor}. 
Although its phase diagram 
has been extensively characterized through Monte Carlo and Molecular Dynamics simulations~\cite{barker1981phase, li2020LJ2D, stepanov2013modelling}, most rheological investigations on this system have been restricted to individual liquid/solid homogeneous phases~\cite{Rizk2022,heyes1986shear,heyes1986molecular,flenner2019viscoelastic,dyre2006colloquium}. 
For instance, 2D LJ liquids exhibit Newtonian regimes followed by pronounced shear thinning associated with structural changes, such as the formation of string-like phases or long-range ordering~\cite{heyes1986shear, heyes1986molecular}. 
However, the viscoelastic response in gas/liquid or gas/solid coexistence regions, where domains with distinct densities and mobilities coexist, remains largely unexplored.
When a fluid exhibits viscoelastic behavior, it is often described by simple models, such as the Maxwell  model that consists of a viscous damper connected in series with a purely elastic spring. 
The shear stress exerted in the linear regime can be expressed as a function of the shear modulus $G^{*}$, which depends on  the storage modulus $G'$ for the elastic response and the loss modulus $G''$ for the viscous dissipation~\cite{grimm2011brownian}. 

The dimensionality of the system plays a decisive role. 
In two-dimensional systems, long-wavelength fluctuations suppress characteristic features of glassy dynamics, such as the intermediate-time plateau in the mean squared displacement~\cite{flenner2019viscoelastic}. This distinguishes viscoelastic relaxation in 2D from that observed in three dimensions, where particle localization is more pronounced~\cite{flenner2019viscoelastic}. 
Consequently, 2D systems may lack a well-defined transient elastic response, unless they reach deep supercooling or exhibit significant structural heterogeneities~\cite{ das2010nonaffine}. 
We recently observed that 2D motility-induced phase separation (MIPS), an out-of-equilibrium analog of a liquid–gas coexistence in an attractive system, fundamentally alters these rheological quantities, showing that the system behaves as a Maxwell-like fluid whose properties are governed by the simultaneous presence of dense and dilute phases~\cite{martin2025motility}.  
This suggests that being at the point of a thermodynamic phase transition in passive systems could similarly modify the rheological properties of systems with coexisting phases.

The present work focuses on analyzing, through two-dimensional Brownian dynamics simulations, how the evolution of  storage/loss  moduli and viscosity  correlate with structural changes in the liquid–gas and solid–gas coexistence regions of passive, thermalized, attractive Lennard-Jones systems. 
We provide a novel perspective by treating rheology not as a property of an isolated phase, but as the collective response of a heterogeneous system at phase coexistence. 
Our objective is to understand the effect of density and temperature on the transition between purely viscous and predominantly elastic behavior across the state diagram, thereby bridging thermodynamic coexistence and nonequilibrium dynamics under shear.

\section{Numerical details and analysis tools}

Our goal is to establish a passive analogue 
of the   Motility Induced Phase Separation~\cite{martin2025motility} observed in Active Brownian Particles. We therefore consider a system of two-dimensional Lennard-Jones particles 
for which the phase diagram has been thoroughly investigated by means of computer simulations~\cite{li2020LJ2D,smit1991vapor}.
This model consists of $N$ disks of diameter 
$\sigma$ interacting via a Lennard-Jones potential of the form
\begin{equation}
    U_{ij}(r_{ij}) = 4 \, \epsilon \, \left[ \left( \frac{\sigma}{r_{ij}}\right)^{12} - \left(  \frac{\sigma}{r_{ij}} \right)^{6} \right] \ ,
\end{equation}
with a cutoff at a distance $r_\text{c}=2.5 \, \sigma$. Here, $r_{ij}(t) = \left|\textbf{r}_i-\textbf{r}_j\right|$ is the  distance between pairs of particles, and $\epsilon$ is the depth of the interaction. 
The  positions of the particles $\textbf{r}_i$ evolve over time according to Brownian dynamics that obey
\begin{equation}
    \gamma \, \frac{d\textbf{r}_i}{dt} = \sum_j \textbf{F}_{ij} + \sqrt{ 2 \gamma k_\text{B} T} \; \boldsymbol{\xi}_i \ ,
\end{equation}
where $\gamma$ is the friction coefficient, 
$k_\text{B}$ the Boltzmann constant, $T$ the temperature. 
The uncorrelated Guassian white noise vector $\boldsymbol{\xi}_i(t)$ satisfies the conditions ($\left\langle \boldsymbol{\xi}_i(t) \right\rangle = \boldsymbol{0}$ and $\left\langle \boldsymbol{\xi}_i(t)\otimes\boldsymbol{\xi}_j(t') \right\rangle = \boldsymbol{1} \delta_{ij} \delta(t-t')$). The conservative interparticle forces $\textbf{F}_{ij}$ are derived from the Lennard-Jones potential. 
The numerical simulations are performed using the   open source molecular dynamics package LAMMPS~\cite{LAMMPS}. 
Throughout the paper, LJ-reduced units are used with $\sigma$ and $\epsilon$   as unit length and energy, respectively.
The friction is set to $\gamma=1$, such that $D=k_\text{B} T$ in all simulations because of the fluctuation-dissipation theorem $\gamma=k_\text{B} T/D$. 
Using  LJ units, our unit of time is $\tau = \sigma^2 \, \gamma/\epsilon =1$. 
The system is composed of the fixed number of particles $N=5000$ in the NVT ensemble with square periodic boundary conditions for different temperatures and  number density $\rho=N/L^2$.

\begin{figure}[h!]
    \centering
    \includegraphics[width=0.75\linewidth]{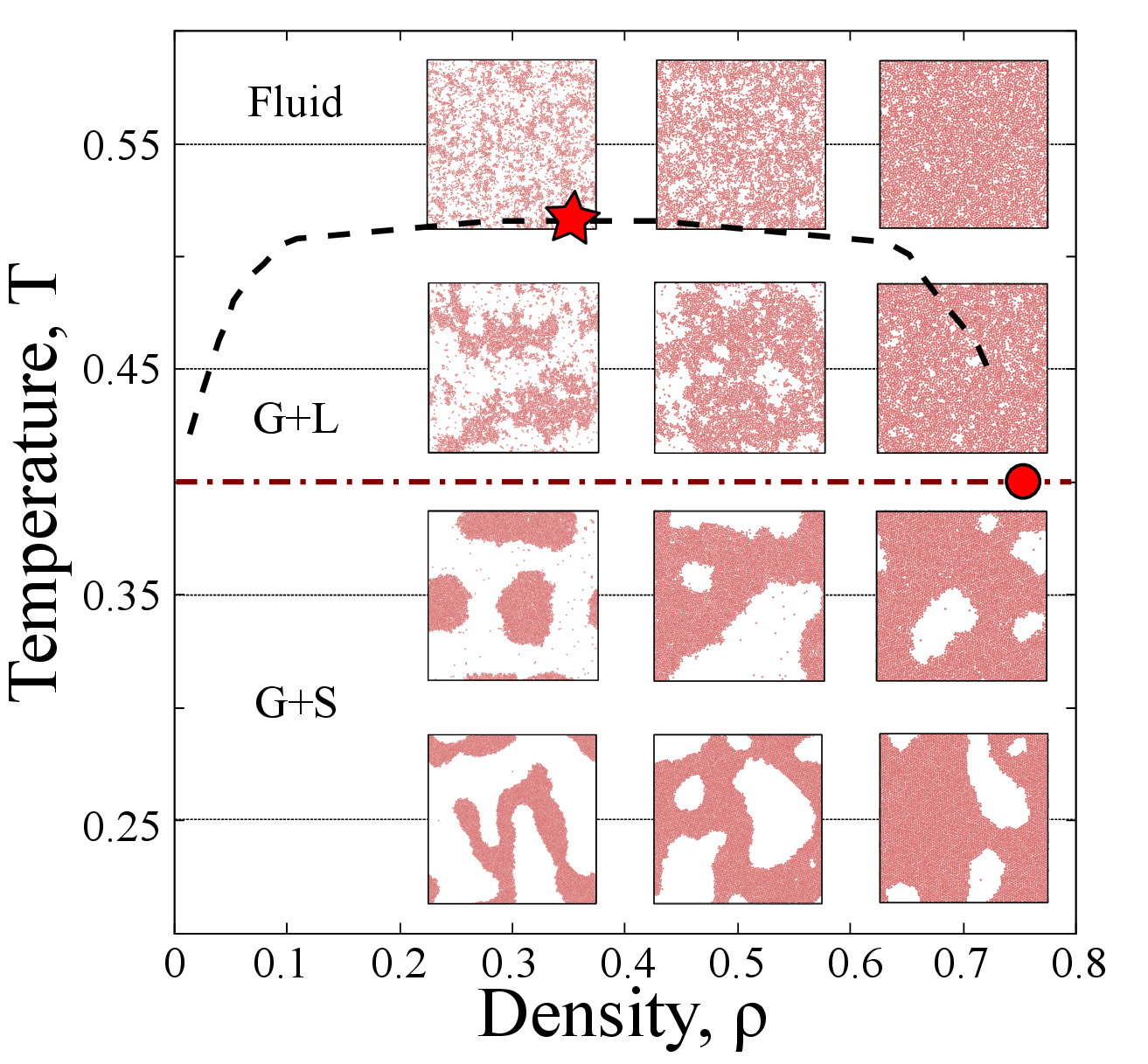}
    \caption{Phase diagram of a two-dimensional Lennard-Jones potential cutoff at $2.5\sigma$. 
    The ashed line corresponds to the liquid/gas coexistence binodal,  the star indicates the location of the critical point $(\rho_\text{c}, \, T_\text{c})=(0.355, \, 0.515)$ and the circle shows the location of the triple point $(\rho_\text{TP}, \, T_\text{TP})=(0.75, \,  0.415)$~\cite{barker1981phase,stepanov2013modelling}.
    G+L labels the gas/liquid coexistence region, while G+S labels the gas/solid coexistence region.
    Snapshots are shown for $\rho \in \left\{ 0.3, 0.5, 0.7 \right\}$ and $T \in \left\{ 0.25, 0.35, 0.45, 0.55 \right\}$.}
    \label{fig:phaseDiagram}
\end{figure}

We focus on a Lennard-Jones suspension truncated at 2.5$\sigma$ whose phase diagram is well established \cite{smit1991vapor,li2020LJ2D}.
As previously reported \cite{smit1991vapor}, a suspension of 2D LJ particles exhibits a binodal line in the T-$\rho$ plane (Fig.~\ref{fig:phaseDiagram}; dashed black line) culminating at the critical point (red star) at $(\rho_\text{c}, \, T_\text{c})=(0.355, \, 0.515)$ ~\cite{barker1981phase,stepanov2013modelling}. 
The system exists as a homogeneous fluid phase above the binodal line~\cite{smit1991vapor} and exhibits gas/liquid coexistence below the binodal line. 
The phase diagram also possesses a triple point $(\rho_\text{TP}, \, T_\text{TP})=(0.75, \,  0.415)$ (Fig.~\ref{fig:phaseDiagram}; red circle), below which (dashed-dotted red line) the gas coexists with the solid phase (gas/solid coexistence). 
Representative snapshots of the system depicting typical configurations in different regions of the phase diagram are shown in Fig.~\ref{fig:phaseDiagram}.
To study the system across the gas/liquid or gas/solid  coexistence regions, the simulations span a broad temperature and density range. 
The systems are first equilibrated using a time-step $\Delta t= 5 \times 10^{-5} \, \tau$ for 
a minimum of $2\times10^4$ steps. Equilibration is monitored by checking the average energy of the system 
as a function of time for all the temperatures and densities. 
Once equilibrated, we launch the production runs for $2 \times 10^9$ steps. 

Rheological measurements are essential for characterizing the response behavior of complex materials. 
Among these, oscillatory shear tests are widely used to extract  elastic and viscous contributions~\cite{martin2025motility, tejedor2023time}. 
However, when systems are in equilibrium, the shear relaxation modulus $G(t)$ in the limit of zero shear,  can be computed directly from the autocorrelation of the  six independent components of the stress tensor, providing a much faster way to compute both storage and loss modulus with respect to direct  oscillatory shear measurements~\cite{boudara2020reptate, tejedor2023time}. 


In 2D, the shear relaxation modulus can be computed using the Green-Kubo relation ~\cite{flenner2019viscoelastic}
\begin{align}
    G(t) =  \frac{A }{k_\text{B} T}  \, \langle \sigma_{xy}(t) \sigma_{xy}(0) \rangle \ , \label{eq:gt2d}
\end{align}
where $A$ is the area of the system. The $xy$ component of the stress tensor due to the pairwise interactions between particles is computed via the virial relation~\cite{flenner2019viscoelastic,heyes1986shear,bayram2023motility} 
\begin{equation}
 \sigma_{xy} = -\frac{1}{2A} \, \sum_{i=1}^N \sum_{j \neq i} \, \frac{\partial U}{\partial r_{ij}}  \, \frac{x_{ij} \, y_{ij}}{r_{ij}} \ ,
\end{equation}
where $x_{ij}$ and $y_{ij}$ are the distances between the centers of two particles $i$ and $j$ along the $x$ and $y$ directions, respectively. 
Once the relaxation modulus $G(t)$ has been computed from  Eq.~\ref{eq:gt2d}, the storage and loss moduli are obtained via the Fourier transform
\begin{equation}
    G^* (\omega) = \omega \,  \int_0^\infty dt \, e^{-i\omega t}  \, G(t) = G' + i \, G'' \ , \label{eq:gpgpp}
\end{equation}
where $G^*$ is the complex shear modulus, whose real part $G'$ corresponds to the storage modulus and whose imaginary part $G''$ corresponds to the loss modulus~\cite{tejedor2023time, martin2025motility}, and $\omega=2\pi/t$ is the frequency. 




\section{Results}



\begin{figure*}[tb]
    \centering
    \includegraphics[width=1\linewidth]{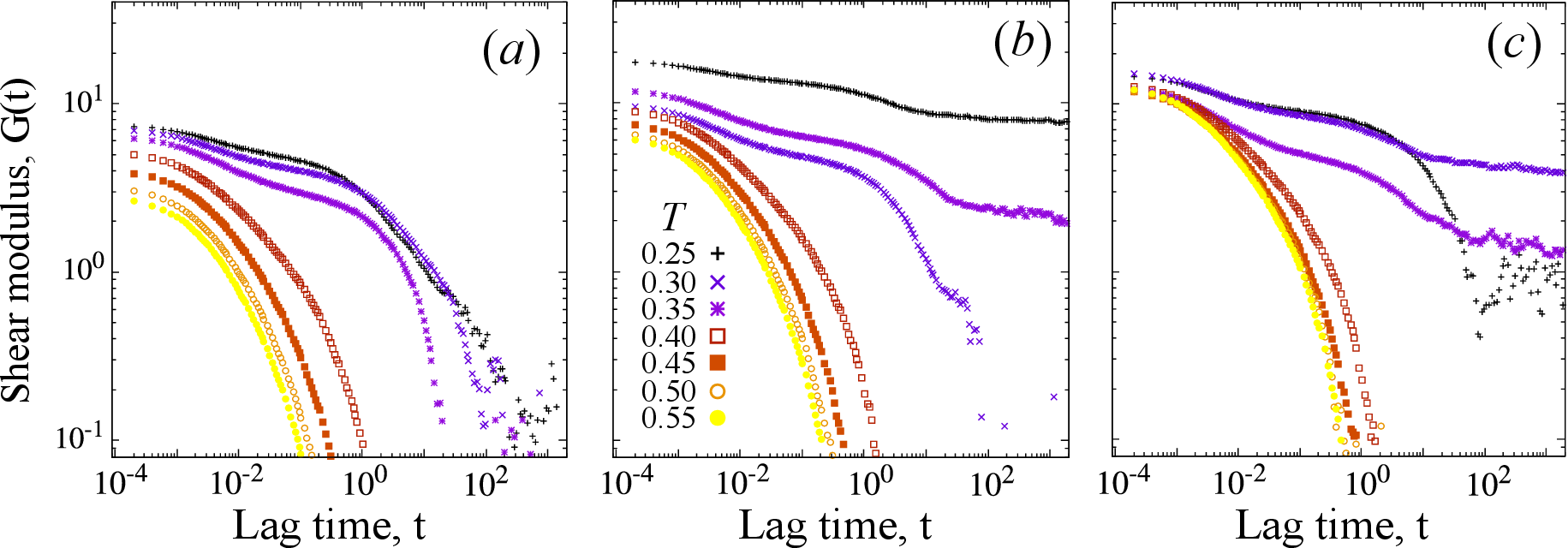}
    \caption{Two dimensional relaxation modulus $G(t)$  computed from Eq.~\ref{eq:gt2d} at different temperatures $T$ for (a) $\rho=0.3$, (b) $\rho=0.5$ and (c) $\rho=0.7$.
    The relaxation modulus decays monotonially as a function of time $t$.}
    \label{fig:gtgpgpp}
\end{figure*}

Figure~\ref{fig:gtgpgpp} shows the shear relaxation modulus $G(t)$ at densities $\rho=0.3$ (Fig.~\ref{fig:gtgpgpp} (a)), $0.5$ (Fig.~\ref{fig:gtgpgpp} (b)) and $0.7$ (Fig.~\ref{fig:gtgpgpp} (c)) and for a range of temperatures between $T=0.55$ (yellow lines/in the gas phase) and $0.25$ (black lines/deep into the gas/solid coexistence region). 
For all densities, 
value of $G(t)$ decreases monotonically with increasing temperature, reflecting the progressive softening of the system as thermal fluctuations overcome interparticle interactions. 

At low density $\rho=0.3$ and higher temperatures $T \geq T_\text{TP}$,  (Fig.~\ref{fig:gtgpgpp}(a)), the relaxation modulus exhibits a single--exponential decay. 
The shape of $G(t)$ is the same for systems above the coexistence binodal ($T=0.55$) as below ($T=0.45$). 
The fact that $G(t)$ decays as a single exponential indicates a single dominant relaxation timescale, which suggests the system behaves as a simple viscoelastic fluid, whose linear response can be effectively described by a Maxwell--like model, despite the gas/liquid binary nature of the system in this regions of the phase diagram.

Upon lowering the temperature, small but systematic deviations from a pure exponential decay become apparent, signaling the emergence of additional relaxation processes beyond a single characteristic time. This happens when $T<T_\text{TP}$, where
the relaxation modulus exhibit two exponential decays: 
One at short times $t\lesssim 2\times10^{-2}$ and another for long times $t\gtrsim10^{-1}$ (Fig.~\ref{fig:gtgpgpp}(a)). 
This double exponential implies the existence of two different relaxation timescales at this temperature and density.  
A similar behavior is observed for intermediate densities $\rho=0.5$ (Fig.~\ref{fig:gtgpgpp}(b)); however, the deviations away from the single exponential become
more pronounced as the temperature decreases into the gas/solid phase. 
This is because, unlike the 
lower density counterpart, at this density the solid percolates across the whole system
(Fig.~\ref{fig:phaseDiagram}; $\rho=0.5$, $T\leq 0.35$ snapshots). 
As a result, $G(t)$ no longer decays to zero at long times. Instead, the decay of  $G(t)$ develops a pronounced long-time tail, indicative of persistent stress correlations and incipient dynamical arrest. 
Finally, at the highest density $\rho=0.7$ (Fig.~\ref{fig:gtgpgpp}(c)), the system displays the same qualitative behavior of the relaxation modulus  $G(t)$ as at intermediate densities.

\begin{figure}[h!]
    \centering
    \includegraphics[width=0.95\linewidth]{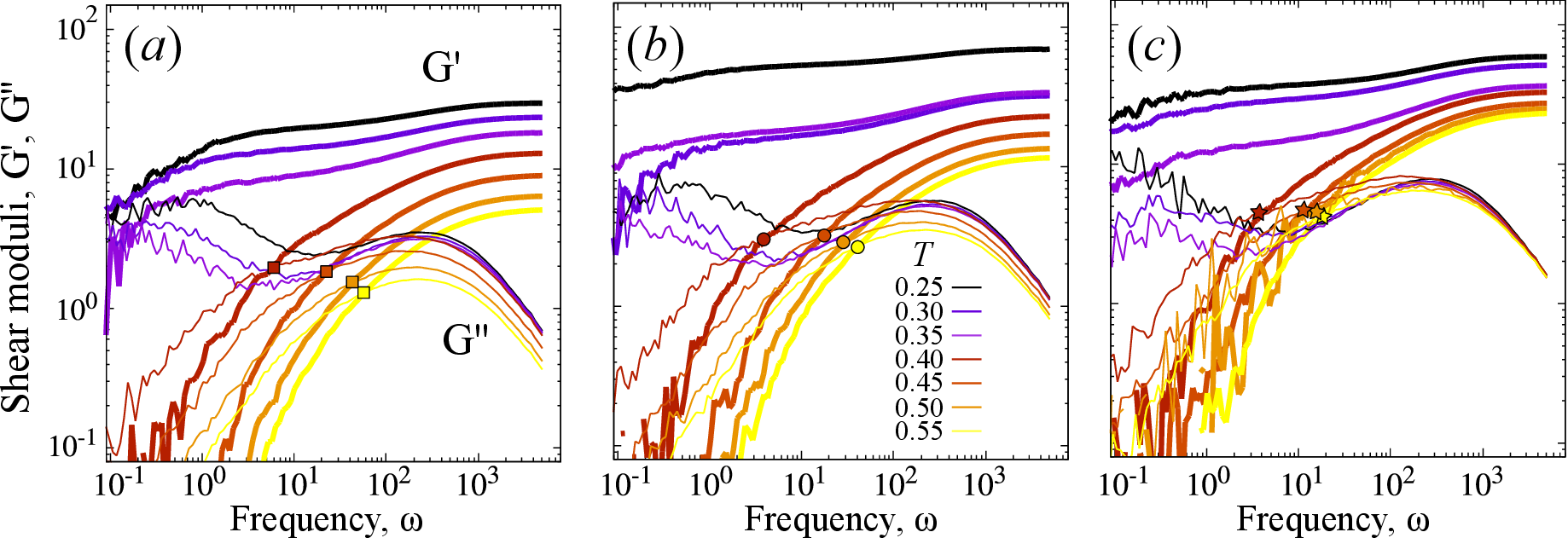}
    \caption{Real $G'(\omega)$ (thick lines) and imaginary $G''(\omega)$ (thin lines) parts of the complex shear modulus at different temperatures, for: (a) $\rho =0.3$, (b) $\rho =0.5$ and (c) $\rho =0.7$. The red, dark orange, orange  and yellow symbols in panels a (squares), b (circles) and c (stars) 
     represent the crossover frequency where $G^{'}$=$G^{''}$
    }
    \label{fig:gp-gpp}
\end{figure}

To further understand the rheological response and relaxation processes of these coexisting phases, Eq.~\ref{eq:gpgpp} can be used to compute the frequency-dependent storage and loss moduli, $G'(\omega)$ and $G''(\omega)$ (Fig.~\ref{fig:gp-gpp}). 
In the high-temperature homogeneous phase ($T=0.5$), $G'(\omega)$ and $G''(\omega)$ exhibit the characteristic scaling of a Maxwell fluid, with $G''(\omega) > G'(\omega)$ at low frequencies and a well-defined crossover frequency $\omega^\dagger$ at which $G'(\omega^\dagger) = G''(\omega^\dagger)$ for all densities (Fig.~\ref{fig:gp-gpp}(a-c)). 
This behavior reflects a predominantly viscous response at long times, which crosses over to an elastic response at short times.
For temperatures and densities where $G(t)$ is characterized by a single exponential decay, $G''(\omega)$ presents a single  local maximum. 

Upon decreasing the temperature and entering the liquid/gas coexistence regime, the system still displays a well-defined crossover frequency at all densities. 
However, the crossover frequency $\omega^\dagger$ shifts to lower values as the temperature is reduced, indicating the progressive slowing of the relaxation dynamics. 
This is consistent with the previous discussion of $G(t)$: systems in the liquid/gas coexistence regime respond in the same qualitative way as a pure LJ fluid with only one relaxation timescale. 
This contrasts with the behavior in the low-temperature regime. Here, the emergence of an additional relaxation time causes a marked deviation from Maxwellian behavior for all densities. Specifically, 
the crossover between $G'(\omega)$ and $G''(\omega)$ disappears from the accessible frequency window, making
the storage modulus $G'(\omega)$ always greater than the loss modulus $G''(\omega)$. 

The dominance of $G'(\omega)$ over $G''(\omega)$ suggests an increasingly elastic response. 
This elastic character is further enhanced at higher densities and lower temperatures, consistent with the formation of larger and denser solid-like structures in the solid–gas coexistence phase.
In the solid/gas coexistence phase ($T<T_\text{TP}$), in addition to $G'(\omega)$ dominating over $G''(\omega)$, the loss modulus $G''$ becomes non-monotonic. 
While $G''(\omega)$ exhibits only a local extremum value at high $T$, it features both a local maximum and a local minimum at low $T$ for all densities. 
This indicates that the system is more viscous at lower frequencies compered with the pure fluid and liquid/gas phases. 
When comparing these mechanical features to those computed for a suspension of active Brownian particles undergoing Motility Induced Phase Separation\cite{martin2025motility}, we observe differences. 
For instance, the cross-over frequency, where $G^{'}$ and $G^{''}$ intersect, shows a non-monotonic behavior with the strength of the active forces for the active system\cite{martin2025motility}, whereas this is not the case for the passive one in its dependence on $T$ (see Fig.\ref{fig:gp-gpp}).
Therefore,  nontrivial rheological features may also arise from  mechanisms associated with structural heterogeneity and coexistence.


To better understand the viscoelastic properties of the system, we also calculate the effective viscosity~\cite{martin2025motility}
\begin{equation}
    \eta(\omega) = \sqrt{\left(\frac{G'}{\omega}\right)^2+\left(\frac{G''}{\omega}\right)^2} \; ,\label{eq:visc}
\end{equation}
as a function of frequency $\omega$. 
Effective viscosity has been previously employed to uncover non-Newtonian behavior in many colloidal systems and complex fluids~\cite{larson1999structure,martin2025motility}.
\begin{figure}[h!]
    \centering
    \includegraphics[width=1\linewidth]{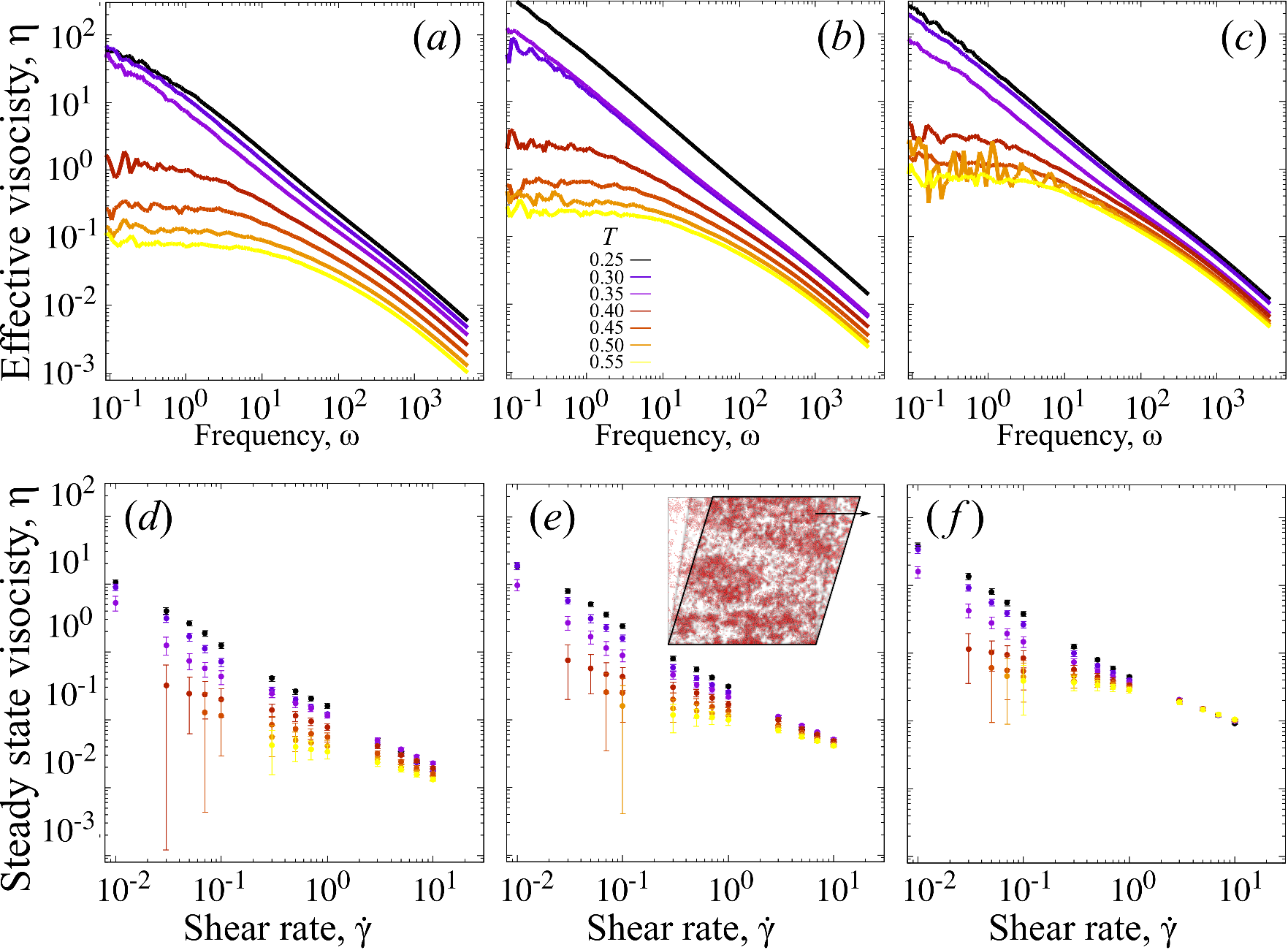}
    \caption{(a-c) Effective viscosity $\eta$ from Eq.~\ref{eq:visc} for a range of frequencies $\omega$ at different temperatures $T$ and density $\rho$. 
    (d-f) The steady-state (st. st.) viscosity as measured by the average out-of-diagonal stress component $\left\langle \sigma_{xy} \right\rangle$ over strain rate $\dot{\gamma}$ in the steady-state after applying a constant strain on the system for different values of temperature and densities. 
    Densities: $\rho=0.3$ (a,d), $\rho=0.5$ (b,e) and $0.7$ (c,f).
    \textbf{Inset}: Snapshot of a typical sheared system at $\rho=0.5$ and T=0.45.
    }
    \label{fig:constant}
\end{figure}
The top panels of Fig.~\ref{fig:constant} (a–c) shows the effective viscosity $\eta(\omega)$ as a function of frequency, computed from Eq.~(\ref{eq:visc}), for several densities and temperatures. 
For temperatures above the triple point, $\eta(\omega)$ exhibits a well-defined low-frequency plateau corresponding to the zero-frequency viscosity of a homogeneous fluid, even though the fluid is a binary coexistance of gas and liquid phases.  
The plateau indicates that, in this regime, the system has sufficient time to fully relax structural stresses, resulting in a predominantly Newtonian response at long times. 
As the temperature is decreased, this low-frequency plateau systematically increases to significantly larger viscosity values and the frequency below which the Newtonian behavior holds shifts to smaller values. 

At higher frequencies, the effective viscosity decreases (Fig.~\ref{fig:constant} (a–c)), which is a clear indicator of shear thinning. 
For these larger frequencies, the imposed deformation occurs faster than the dominant structural relaxation times, preventing the system from fully reorganizing and leading to a reduced effective viscosity. 
This is true for all temperatures and densities. 
Indeed, below the triple point temperature, no Newtonian plateau is observed at the computationally accessible frequencies. 
The disappearance of the plateau is a direct rheological signature of phase coexistence and of the increasing solid-like character of the system. 
Because it does not plateau, the effective viscosity continues to rise as the frequency is decreased. 
This increased low-frequency viscosity reflects the emergence of long-lived stress correlations associated with the presence of extended solid domains, which strongly hinder structural relaxation and effectively enhance the resistance to flow. 

Qualitatively, the same behavior is seen for all densities (Fig.~\ref{fig:constant} (a–c)), which is expected given the flattened shape of the phase diagram with respect to $\rho$. 
Overall, across the range of densities and temperatures explored within the coexistence regions, the system consistently exhibits a pronounced tendency toward shear thinning. 
This behavior is consistent with theoretical and numerical predictions for Lennard-Jones fluids, where the imposed shear rate competes with thermal motion, leading to a redirection of particle trajectories and a reduction of the effective viscosity~\cite{heyes1986shear}.


To fully characterize the rheological response, the dependence of the viscosity on frequency is compared with nonlinear constant-shear measurements. In the linear regime, for small shear rates $\dot{\gamma}$, the steady-state shear stress $\langle \sigma_{xy} \rangle$ (Fig.~\ref{fig:constant}; inset) is evaluated for different shear rates $\dot{\gamma}$ and the steady-state viscosity is given by 
\begin{equation}
    \eta(\dot{\gamma}) = \frac{ \left\langle \sigma_{xy} \right\rangle }{ \dot{\gamma} } \ . 
\end{equation} 
Beyond the low-frequency Newtonian plateau, where the viscosity is independent of the shear rate, the system response follows a power-law relationship  given by the Herschel–Bulkley model\cite{saramito2016complex,heyes1986shear, bayram2023motility, martin2025motility}
\begin{equation}
    \sigma_{xy} =\sigma_0 +k \,  \dot{\gamma}^n \ , 
    \label{eq:model}
\end{equation}
where $\sigma_0$ represents the yield stress, $k$ is a fit parameter and the value of the stress exponent $n$ classifies the fluid behavior. 
An exponent of $n < 1$ indicates shear thinning~\cite{heyes1986shear,vazquez2016shear}, while $n > 1$ represents shear thickening~\cite{wyart2014discontinuous,lin2015hydrodynamic} and $n = 1$ corresponds to a purely Newtonian fluid. 
Homogeneous Lennard-Jones (LJ) fluids have been shown to undergo a marked transition from a Newtonian regime at low shear rates to a pronounced shear-thinning regime at higher deformation rates~\cite{heyes1986shear}. 

Analyzing how $n$ evolves across the phase diagram establishes a correlation between the thermodynamic coexistence state (liquid/gas or solid/gas) and the emergence of non-Newtonian regimes.
In phase-separating systems, 
the exponent $n$ serves as a sensitive indicator of the system’s rheological responses, with values of $n < 1$ signaling persistent shear-thinning behavior~\cite{martin2025motility}, typically associated with microscopic structural reorganization. 
In this regime, particles organize into anisotropic structures or string-like configurations that facilitate flow and reduce viscous resistance~\cite{heyes1986shear, bayram2023motility, martin2025motility}.

For high temperatures, $\eta(\dot{\gamma})$ exhibits a plateau at low $\dot{\gamma}$ for all densities (Fig.~\ref{fig:constant} (d–f)). 
This suggests $n=1$, indicating a purely Newtonian fluid, even inside the gas/liquid coexistence region. 
As the shear rate is increased, $\eta(\dot{\gamma})$ begins to decrease, indicating a shear thinning fluid with $n<1$, in agreement with the effective viscosity (Fig.~\ref{fig:constant}(a-c)). 
As the temperature is lowered ($T<T_\text{TP}$), the Herschel-Buckley stress exponent is less than unity ($n<1$) for all shear rates explored here, demonstrating shear thinning for all systems below the triple point, \textit{i.e.} all gas/solid coexistence phases. 

The qualitative agreement between the frequency-dependent viscosity $\eta(\omega)$ at large frequency $\omega$ (Fig.~\ref{fig:constant} (a-c)) and the steady-state viscosity $\eta(\dot{\gamma})$  at large shear rates (Fig.~\ref{fig:constant} (d-f)) demonstrates the consistency between linear response rheology and nonlinear constant-shear measurements. 
Both approaches probe regimes in which the externally imposed deformation exceeds internal relaxation processes, giving rise to similar shear-thinning phenomenology. 
Importantly, this correspondence holds across the different coexistence regimes explored, indicating that shear thinning is a robust feature of the Lennard-Jones system in two dimensions when phase heterogeneity is present.

\section{Conclusions}

In this work, we have investigated the rheological properties of a two dimensional Lennard-Jones system across liquid/gas and solid/gas coexistence regions by combining linear-response viscoelastic measurements with nonlinear steady-shear simulations. 
By explicitly accounting for phase coexistence, these results provide a more complete picture of the rheological properties of simple liquids by explicitly accounting for their behavior in the presence of coexisting phases.


This work has shown that the relaxation dynamics undergo a qualitative change upon entering coexistence regions. 
At high temperatures, the system exhibits a Maxwell-like viscoelastic response characterized by a single dominant relaxation time. 
This is true for pure fluids at temperatures above the liquid/gas coexistence binodal. 
Perhaps more unexpectedly, this is also true below the binodal for heterogeneous systems composed of a coexistence of gas and liquid phases. 
Further lowering the temperature below the triple point leads to the emergence of multiple relaxation timescales, long-lived stress correlations, and a progressive enhancement of elastic behavior. 
This is because the system has entered a heterogeneous solid/gas coexistence regime, where extended solid domains strongly hinder stress relaxation and give rise to large effective viscosities.

The frequency-dependent moduli and the effective viscosity provide a clear connection between structural heterogeneity and rheological response. 
In particular, the monotonic decrease of the crossover frequency $\omega^\dagger$ with decreasing temperature reflects the continuous slowing of stress relaxation as the system approaches gas/solid coexistence. 
At the same time, both oscillatory and steady-shear measurements consistently indicate the presence of shear-thinning behavior across all three regimes, gas/solid coexistence, gas/liquid coexistence and single phase fluid. 
This emphasizes the robustness of this non-Newtonian response in two-dimensional attractive systems.

Although the present study focuses on a passive system, this work was partly motivated by recent findings in active matter, where phase separation has been shown to strongly modify viscoelastic properties~\cite{martin2025motility}. 
A comparison with active systems~\cite{martin2025motility} underscores that phase coexistence alone is sufficient to generate complex rheological behavior in equilibrium systems, even in the absence of self-propulsion. 
Although there are fundamental differences between the system studied in this paper and its active counterpart, certain rheological features previously attributed to activity may instead arise from more general mechanisms associated with structural heterogeneity and coexistence.

Overall, these results establish a direct link between thermodynamic phase behavior and rheological response in two-dimensional Lennard-Jones systems. 
They provide a unified framework for understanding how density, temperature, and phase coexistence control the transition from viscous to elastic-dominated behavior under shear, and offer a reference point for future studies of passive and active heterogeneous materials.

\ack{We acknowledge useful discussions with James Richards.}

\funding{This research has received funding from the European Research Council (ERC) under the European Union’s Horizon 2020 research and innovation program (Grant Agreement No. 851196). A.C. acknowledges financial support from the National Science Foundation
under Grant No. DMR-2321925.} 
\textcolor{black}{C.V. acknowledges fundings IHRC22/00002 and Proyecto PID2022-140407NB-C21 funded by
MCIN/AEI /10.13039/501100011033 and FEDER, UE.}

\roles{All authors contributed to the conception of the work. They also were involved in drafting or revising the article critically for intellectual content. All approved the final version.}

\data{The data that support the findings of this study are available
from the corresponding author upon reasonable request.}





\providecommand{\newblock}{}

\end{document}